# Observation of the Quantum Spin Hall Effect up to 100 Kelvin in a Monolayer Crystal

**Authors:** Sanfeng Wu[1, #, *], Valla Fatemi[1, #, *], Quinn D. Gibson[2], Kenji Watanabe[3], Takashi Taniguchi[3], Robert J. Cava[2], and Pablo Jarillo-Herrero[1, *]

**Affiliations:**
[1]Department of Physics, Massachusetts Institute of Technology, Cambridge, Massachusetts 02139, USA
[2]Department of Chemistry, Princeton University, Princeton, New Jersey 08544, USA
[3]Advanced Materials Laboratory, National Institute for Materials Science, 1-1 Namiki, Tsukuba 305-0044, Japan
[#] These authors contributed equally to this work.
[*]Corresponding Email: swu02@mit.edu; vfatemi@mit.edu; pjarillo@mit.edu

**Abstract**: The field of topological insulators (TI) was sparked by the prediction of the quantum spin Hall effect (QSHE) in time reversal invariant systems, such as spin-orbit coupled monolayer graphene. Ever since, a variety of monolayer crystals have been proposed as two-dimensional (2D) TIs exhibiting the QSHE, possibly even at high temperatures. However, conclusive evidence for a monolayer QSHE is still lacking, and systems based on semiconductor heterostructures operate at temperatures close to liquid helium. Here we report the observation of the QSHE in monolayer $WTe_2$ at temperatures up to 100 Kelvin. The monolayer exhibits the hallmark quantized transport conductance, ~ $e^2/h$ per edge, in the short edge limit. Moreover, a magnetic field suppresses the conductance, and the observed Zeeman-type gap indicates the existence of a Kramers degenerate point, demonstrating the importance of time reversal symmetry for protection from elastic backscattering. Our results establish the high-temperature QSHE and open a new realm for the discovery of topological phases based on 2D crystals.

**Text** – A time-reversal (TR) invariant topological insulator (TI) in two dimensions (2D), also known as a quantum spin Hall (QSH) insulator, can be identified by its unique helical edge modes (*1–4*). So far, evidence for the helical edge mode in 2D TIs, particularly quantized transport, has been limited to very low temperatures (i.e. near liquid helium temperature) in HgTe and InAs/GaSb quantum wells (*5, 6*). In the search for high temperature TIs, substantial efforts have focused on a variety of atomically thin materials (*7–14*), which have the additional promise of advancing the field of topological physics using the tools developed for 2D crystals. However, experimental observation of the quantum spin Hall effect (QSHE) in monolayer systems is challenging, often due to structural or chemical instabilities (*9, 15–17*). Indications of a high temperature QSH phase in bulk-attached bismuth bilayers have been reported (*7, 18, 19*), but a conclusive demonstration is still lacking.

Among the proposals for atomically thin TIs are monolayer transition metal dichalcogenides (TMDs), materials that are either 2D semiconductors or semimetals depending on their structural phase (*9*). Calculations suggest that an inverted band gap can develop in 1T' TMD monolayers, resulting in a nontrivial $Z_2$ topological phase (*9, 20*). Recent experiments have shown promising results (*12–14*), including that 1T' monolayer $WTe_2$ exhibits a ground state with an insulating interior and conducting edges associated with a zero-bias anomaly (*12*), distinct from its multilayer counterparts (*12*, *21*). However, the QSHE, the hallmark of a 2D TI, has yet to be observed, and hence its topological nature is far from conclusive. Here we observe the QSHE in WTe$_2$ monolayers and identify this 2D material as an atomically layered TI with conductance ~ $e^2/h$ per edge at high-temperatures.

QSH transport through a 2D TR-invariant TI should exhibit the following characteristics: (a) helical edge modes, characterized by an edge conductance that is approximately the quantum value of $e^2/h$ per edge (*5*); (b) saturation to the conductance quantum in the short edge limit (*22*); and (c) suppression of conductance quantization upon application of a magnetic field, due to the loss of protection by TR symmetry (*5*, *23*, *24*). Signatures of a Zeeman gap should be seen if the Kramer's degeneracy (Dirac point) is located inside the bulk bandgap. To date, simultaneous observation of the above criteria in existing 2D TI systems is still lacking (*5*, *6*, *22*, *23*, *25*), prompting the search for new QSH materials.

To check the above criteria in monolayer 1T'-WTe$_2$, we fabricated devices with the structure depicted in Fig. 1A (see also Fig. S1, S2 and Materials & Methods). The goal of the design has three objectives: (1) ensure an atomically flat, chemically protected channel (no flake bending or exposure) by fully encapsulating the flake with hexagonal boron nitride (*15*, *21*); (2) minimize the effect of contact resistance; and (3) enable a length-dependence study on a single device. Our devices generally consist of eight contact electrodes, a top graphite gate, and a series of in-channel local bottom gates with width $L_c$ varying from 50 nm to 900 nm. The monolayer flakes are carefully selected to have a long strip shape, typically a few μm wide and about ten μm long (Table S1). Fig. 1B shows a typical measurement of the four-probe conductance (in Device 1) across all the local gates (~ 8 μm long) as a function of top gate voltage, $V_{tg}$. A finite conductance plateau develops around $V_{tg} = 0$ V. This characteristic feature for monolayer WTe$_2$ is due to conduction along the edges (*12*). The measured value is highly sensitive to (typically poor) contact properties (*12*), which prevents observation of the intrinsic edge conductance. We overcome this obstacle in our devices through selective doping of the flake using a combination of global top and local bottom gates. A short transport channel with length $L_c$ can be selectively defined by a local gate voltage $V_c$, while the rest of the flake is highly doped by $V_{tg}$ to secure good contact to the electrodes (see Fig. S3 for dI/dV characteristics). Figure 1C maps out the resistance $R$ in the same device as a function of $V_{tg}$ and $V_c$ (for a local gate with $L_c$ =100 nm). The step structure indicates a transition from a bulk-metallic state (doped) to a bulk-insulating state (undoped) within the locally gated region. We define the offset resistance, $\Delta R = R(V_c) - R(V_c = -1V)$, as the resistance change from the value in the highly doped limit ($V_c = -1V$ in this case). Figure 1D shows a $\Delta R$ trace (red curve) extracted from Fig 1C (dashed white line in Fig. 1C), where $V_{tg}$ is fixed at 3.5 V. The average value of $\Delta R$ at the plateau, which measures the step height, saturates when $V_{tg}$ is high enough (Fig. 1D inset and Fig. S4-7).

This saturated value, $\Delta R_s$, thus measures the resistance of the undoped channel, which can only originate from the edges, because the bulk is insulating (*12*, *14*, *13*). Notably, $\Delta R_s$ is approximately equal to $h/2e^2$ for both this 100 nm channel and the 60 nm and 70 nm channels on Device 2 (Fig. 1D). Fluctuations in the range of few kΩ, which may originate from residual disorder or correlation effects (*12*, *26*, *27*), are visible but decrease substantially above 4K. Given that the sample has two edges, the observed conductance per edge is therefore ~ $e^2/h$, pointing to helical edge modes as the source of the conductance (*5, 6*). In order to confirm this scenario, one must rule out the possibility of trivial diffusive edge modes that happen to exhibit the quantized conductance value for some particular length (*22*). We thus performed a length dependence study utilizing a series of local gates with different $L_c$. Detailed analysis of measurements from representative devices and gates at ~ 4 K can be found in Figs. S3-5. In Fig. 2 we summarize the data by plotting the undoped channel resistance, $\Delta R_s$, as a function of $L_c$. For long edges the resistance generally decreases with decreasing length, which is arguably captured by a linear trend. The behavior, however, clearly deviates from the trend when $L_c$ is reduced to 100 nm or less, where the resistance saturates to a value close to $h/2e^2$. Such behavior is present in all three devices that enter this short-length regime, independent of the width of the monolayer flake (varying from 1 to 4 µm). These observations reveal the intrinsic conductance as $e^2/h$ per edge as per the abovementioned criteria (a) and (b) for the QSHE.

To check criterion (c), regarding TR symmetry protection from elastic scattering, we performed magneto-conductance measurements. The data taken from the 100 nm long channel in Device 1 in the QSHE regime (i.e. gate range on plateau) is shown in Fig. 3. We define $G_s$ as $1/\Delta R_s$, which measures the conductance of the edges in the short channel limit. $G_s$ is plotted as a function of $V_c$ in Fig. 3A for a series of magnetic fields $B$ applied perpendicular to the monolayer at 1.6 K. $G_s$ decreases significantly once $B$ is turned on, in contrast to the bulk state which is hardly affected (Fig. S8). For all $V_c$, $G_s$ decreases rapidly for low magnetic fields ($B < 2T$). After this initial stage, two types of behavior are observed, depending on $V_c$, as shown in Fig. 3B. When $V_c$ is near -6.44V, $G_s$ decreases exponentially, without saturation up to 8 T. For other values of $V_c$, $G_s$ saturates at high $B$. These behaviors are significantly different from the previous observations for resistive channels (*12*).

Both types of behavior can be understood in the context of the QSHE. The 1D edge state of the QSH phase consists of two species: left and right movers associated with opposite spin polarization. The two linearly dispersing bands cross at the Kramers degeneracy point (Fig 3B, panel a). Magnetic fields applied nonparallel to the spin polarization are expected to open an energy gap at the Kramers point due to the Zeeman effect (*28*). For a homogenous chemical potential close to the degeneracy point (Fig 3B, panel b), one would expect an exponential decay of the conductance without saturation. To reveal the existence of the gap, we performed temperature dependence measurements of the magneto-conductance at $V_c = -6.44V$. The exponential decay of $G_s$ persists up to high temperatures (measured up to 34 K, inset of Fig. 3C). Moreover, all the curves collapse onto a single universal trend when renormalized by plotting the dimensionless values $-\log(G_s/G_0)$ vs $\mu_B B/k_B T$ (Fig. 3C), where $G_0$ is the zero-field conductance, $\mu_B$ is the Bohr magneton, $k_B$ is the Boltzmann constant, and $T$ is the temperature. The slope of the trend yields an effective g-factor ~ 4.8 for the out-of-plane field in this device (i.e. the device conductance obeys $G_s = G_0 \exp(-g\mu_B B/2k_B T)$). This observation confirms a Zeeman-type gap opening in the edge bands.

If the Fermi energy at the edge is gated away from the Kramers point (Fig 3B, panel c), the Zeeman gap will not be directly observed, and the magneto-conductance should be determined by the scattering mechanisms at the edge allowed by the TR symmetry breaking. For example, in our devices the presence of local charge puddles can be natural. According to theoretical calculations, the edge conductance will be reduced to $\alpha e^2/h$, where $\alpha$ is determined by the microscopic details of the edge (*24*, *29*). Calculations show that at high magnetic fields an individual puddle can reduce transmission along an edge by 50% (*24*, *30*), leading to a saturated α determined by the distribution of the puddles along the edges. We find the conductance saturation is consistent with this picture (Fig. S9). In addition to vertical magnetic fields, we have also found significantly reduced edge conductance when in-plane magnetic field is applied (Fig. S10). We expect that both in- and out-of-plane magnetic fields will suppress the conductance because breaking time reversal symmetry removes protection of the edge conduction and the edge mode spin polarization axis is not necessarily normal or parallel to the layer due to the lack of out-of-plane mirror symmetry in the monolayer. The exact spin polarization axis may be influenced by multiple factors, such as the direction of the crystallographic edge and the existence of displacement electric fields. The irregular edge of the exfoliated monolayer makes the situation more complex. Overall, the magneto-conductance behavior reveals the expected necessity of TR symmetry for the QSHE to be exhibited, and thus confirms criterion (c). Therefore, the QSHE is indeed observed in monolayer WTe$_2$.

Remarkably, the distinctive conductance value survives up to high temperatures. Figure 4A plots the temperature dependence of $G_s$ at different $V_c$ in the QSHE regime; $G_s$ stays approximately constant and close to $2e^2/h$ up to 100 K, indicating that the conductance is dominated by the QSHE up to this temperature. In terms of $\Delta R$, the resistance plateau starts to drop at around 100 K (Fig. 4A inset). We note that the it is not obvious a priori what the temperature dependence of the QSH edge conductance should be, and some proposed mechanisms indicate weak (*31*) or even negative temperature dependence (*26*). Above this temperature, the channel conductance increases rapidly with temperature, indicating the activation of bulk conduction channels. To reveal the transition more clearly, in Fig. 4B we plot the temperature dependence of the resistance $R$ of the whole flake (i.e. entire length, which consists of locally gated region in series with the rest of the flake) when the Fermi energy in the local channel is placed in the metallic regime ($V_c$ = -1V) and the QSH regime ($V_c$ < -5.3V). A clear kink at 100 K can be seen in the QSH regime. The difference between the two curves yields the channel resistance which drops above the transition temperature.

This high temperature QSHE is consistent with the prediction of a large inverted bandgap (~100 meV) in monolayer WTe$_2$ (*20*) as well as recent experiments that observe a ~45 meV bulk bandgap in spectroscopy (*13*, *14*) and a similar onset temperature for bulk conduction (*12*). We suspect the 100 K transition temperature may not be an intrinsic limit. Improvements in device quality may enable observation of the QSHE at even higher temperatures and for longer edges.

Our observations have confirmed the nontrivial TR invariant topological phase in monolayer 1T'-WTe$_2$ and have demonstrated the QSHE at high temperatures for the first time in an isolated 2D monolayer device. The exploration of 2D topological physics and device performance above liquid nitrogen temperatures has therefore become possible. Distinct from quantum well systems, the exposed nature of isolated monolayers may allow to engineer topological phases in unprecedented ways. In particular, WTe$_2$ can be readily combined with other 2D materials to form novel van der Waals heterostructures, a promising platform for studying the proximity effect between a QSH system and superconductors or magnets (*3*, *4*) at the atomic scale.

**References and Notes:**

1. C. L. Kane, E. J. Mele, Quantum Spin Hall Effect in Graphene. *Phys. Rev. Lett.* **95**, 226801 (2005).

2. B. A. Bernevig, S.-C. Zhang, Quantum Spin Hall Effect. *Phys. Rev. Lett.* **96**, 106802 (2006).

3. M. Z. Hasan, C. L. Kane, Colloquium: Topological insulators. *Rev. Mod. Phys.* **82**, 3045–3067 (2010).

4. X.-L. Qi, S.-C. Zhang, Topological insulators and superconductors. *Rev. Mod. Phys.* **83**, 1057–1110 (2011).

5. M. König *et al.*, Quantum Spin Hall Insulator State in HgTe Quantum Wells. *Science*. **318**, 766–770 (2007).

6. I. Knez, R.-R. Du, G. Sullivan, Evidence for Helical Edge Modes in Inverted InAs/GaSb Quantum Wells. *Phys. Rev. Lett.* **107**, 136603 (2011).

7. S. Murakami, Quantum Spin Hall Effect and Enhanced Magnetic Response by Spin-Orbit Coupling. *Phys. Rev. Lett.* **97**, 236805 (2006).

8. Y. Xu *et al.*, Large-Gap Quantum Spin Hall Insulators in Tin Films. *Phys. Rev. Lett.* **111**, 136804 (2013).

9. X. Qian, J. Liu, L. Fu, J. Li, Quantum spin Hall effect in two-dimensional transition metal dichalcogenides. *Science*. **346**, 1344–1347 (2014).

10. J.-J. Zhou, W. Feng, C.-C. Liu, S. Guan, Y. Yao, Large-Gap Quantum Spin Hall Insulator in Single Layer Bismuth Monobromide Bi4Br4. *Nano Lett.* **14**, 4767–4771 (2014).

11. S. Li *et al.*, Robust Room-Temperature Quantum Spin Hall Effect in Methyl-functionalized InBi honeycomb film. *Sci. Rep.* **6**, 23242 (2016).

12. Z. Fei *et al.*, Edge conduction in monolayer WTe2. *Nat. Phys.* **advance online publication** (2017), doi:10.1038/nphys4091.

13. S. Tang *et al.*, Quantum spin Hall state in monolayer 1T'-WTe2. *Nat. Phys.* **13**, 683–687 (2017).

14. Z.-Y. Jia *et al.*, Direct visualization of a two-dimensional topological insulator in the single-layer 1T'-WTe2. *Phys. Rev. B*. **96**, 041108 (2017).

15. Y. Cao *et al.*, Quality Heterostructures from Two-Dimensional Crystals Unstable in Air by Their Assembly in Inert Atmosphere. *Nano Lett.* **15**, 4914–4921 (2015).

16. L. Wang *et al.*, Tuning magnetotransport in a compensated semimetal at the atomic scale. *Nat. Commun.* **6**, 8892 (2015).

17. F. Ye *et al.*, Environmental Instability and Degradation of Single- and Few-Layer WTe2 Nanosheets in Ambient Conditions. *Small*. **12**, 5802–5808 (2016).

18. C. Sabater *et al.*, Topologically Protected Quantum Transport in Locally Exfoliated Bismuth at Room Temperature. *Phys. Rev. Lett.* **110**, 176802 (2013).

19. I. K. Drozdov *et al.*, One-dimensional topological edge states of bismuth bilayers. *Nat. Phys.* **10**, 664–669 (2014).

20. F. Zheng *et al.*, On the Quantum Spin Hall Gap of Monolayer 1T′-WTe2. *Adv. Mater.* **28**, 4845–4851 (2016).

21. V. Fatemi *et al.*, Magnetoresistance and quantum oscillations of an electrostatically tuned semimetal-to-metal transition in ultrathin WTe2. *Phys. Rev. B*. **95**, 041410 (2017).

22. F. Nichele *et al.*, Edge transport in the trivial phase of InAs/GaSb. *New J. Phys.* **18**, 083005 (2016).

23. E. Y. Ma *et al.*, Unexpected edge conduction in mercury telluride quantum wells under broken time-reversal symmetry. *Nat. Commun.* **6**, 7252 (2015).

24. S. Essert, K. Richter, Magnetotransport in disordered two-dimensional topological insulators: signatures of charge puddles. *2D Mater.* **2**, 024005 (2015).

25. L. Du, I. Knez, G. Sullivan, R.-R. Du, Robust Helical Edge Transport in Gated InAs/GaSb Bilayers. *Phys. Rev. Lett.* **114**, 096802 (2015).

26. J. Maciejko *et al.*, Kondo Effect in the Helical Edge Liquid of the Quantum Spin Hall State. *Phys. Rev. Lett.* **102**, 256803 (2009).

27. T. Li *et al.*, Observation of a Helical Luttinger Liquid in InAs/GaSb Quantum Spin Hall Edges. *Phys. Rev. Lett.* **115**, 136804 (2015).

28. M. Konig *et al.*, The Quantum Spin Hall Effect: Theory and Experiment. *J. Phys. Soc. Jpn.* **77**, 031007 (2008).

29. J. Maciejko, X.-L. Qi, S.-C. Zhang, Magnetoconductance of the quantum spin Hall state. *Phys. Rev. B*. **82**, 155310 (2010).

30. A. Roth *et al.*, Nonlocal Transport in the Quantum Spin Hall State. *Science*. **325**, 294–297 (2009).

31. J. I. Väyrynen, M. Goldstein, Y. Gefen, L. I. Glazman, Resistance of helical edges formed in a semiconductor heterostructure. *Phys. Rev. B*. **90**, 115309 (2014).

32. M. N. Ali *et al.*, Correlation of crystal quality and extreme magnetoresistance of WTe 2. *EPL Europhys. Lett.* **110**, 67002 (2015).

33. M. Kim *et al.*, Determination of the thickness and orientation of few-layer tungsten ditelluride using polarized Raman spectroscopy. *2D Mater.* **3**, 034004 (2016).

**Acknowledgments:**
We thank Liang Fu and Xiaofeng Qian for helpful discussions. This work was partly supported through AFOSR Grant No. FA9550-16-1-0382 as well as the Gordon and Betty Moore Foundation's EPiQS Initiative through Grant No. GBMF4541 to P. J-H. Device nanofabrication was partly supported by the Center for Excitonics, an Energy Frontier Research Center funded by the DOE, Basic Energy Sciences Office, under Award No. DE-SC0001088. This work made use of the Materials Research Science and Engineering Center's Shared Experimental Facilities supported by NSF under Award No. DMR-0819762. Sample fabrication was performed partly at the Harvard Center for Nanoscale Science supported by the NSF under Grant No. ECS-0335765. S.W. acknowledges the support of the MIT Pappalardo Fellowship in Physics. The WTe2 crystal growth performed at Princeton University was supported by an NSF MRSEC grant, DMR-1420541. Growth of hexagonal boron nitride crystals was supported by the Elemental Strategy Initiative conducted by the MEXT, Japan and JSPS KAKENHI Grant Numbers JP15K21722 and JP25106006.

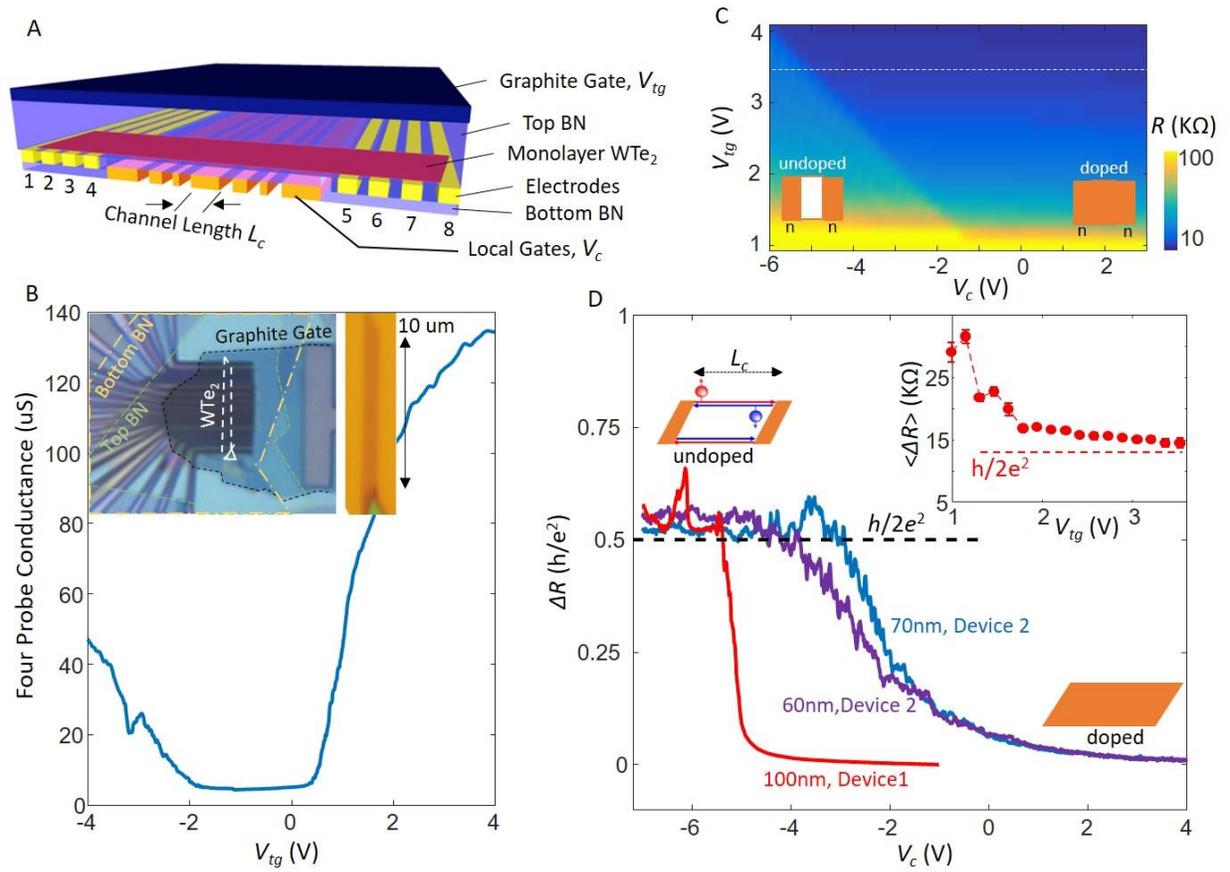

**Fig. 1**. **Device structure and resistance near $h/2e^2$**. (A) Schematic of the device structure. (B) Four probe conductance measurement at 4 K of Device 1 as a function of $V_{tg}$ across all the local gates, which are floating. Inset: the optical image of Device 1 (left) and the corresponding monolayer WTe$_2$ flake before fabrication (right). (C) Color map of the flake resistance tuned by $V_{tg}$ and the 100nm-wide local gate $V_c$ at 4K. Two regions are separated by a step in the resistance distinguishing the doped and undoped local channel, as depicted by the inset schematics. (D) $\Delta R$ versus $V_c$ for the 100 nm wide gate on Device 1 at $V_{tg} = 3.5$V, and the 60 and 70 nm wide gates on Device 2 at $V_{tg} = 4.1$V (taken at 5 K). For clarity, the two curves from Device 2 are offset by +3 V along the x axis. Inset: the average step height $\langle\Delta R\rangle$, extracted from (C), as a function of $V_{tg}$, showing a clear saturation towards $h/2e^2$ for large $V_{tg}$.

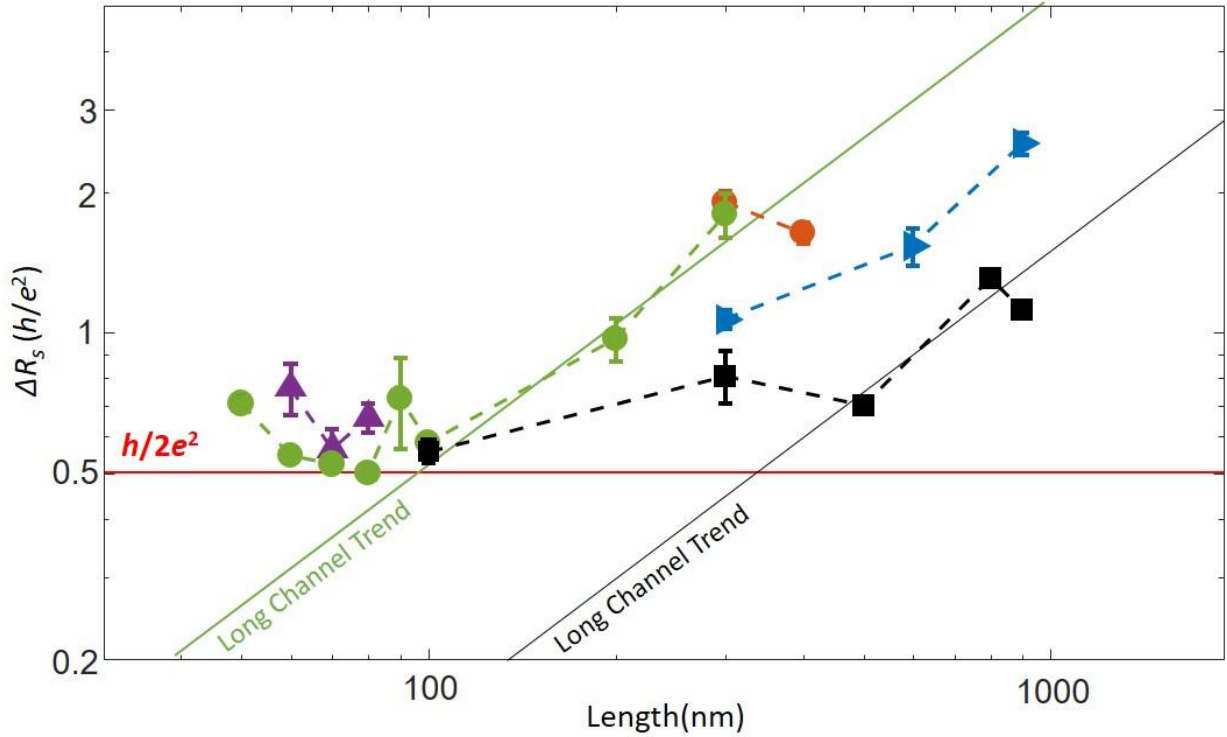

**Fig. 2. Length dependence of the undoped channel resistance**. Data taken at 4 K from 5 different devices (Table S1), each denoted by a different color and symbol. The device numbers and associated colors are: 1, Black; 2, green; 3, purple; 4, red; 5, blue. The $\Delta R_s$ values approach a minimum of $h/2e^2$ in the short-channel limit, confirming a total conductance of $2e^2/h$ for the undoped channel, i.e. a conductance of $e^2/h$ per edge in the device, in agreement with QSHE. Detailed analysis of raw data can be found in Fig. S4-7.

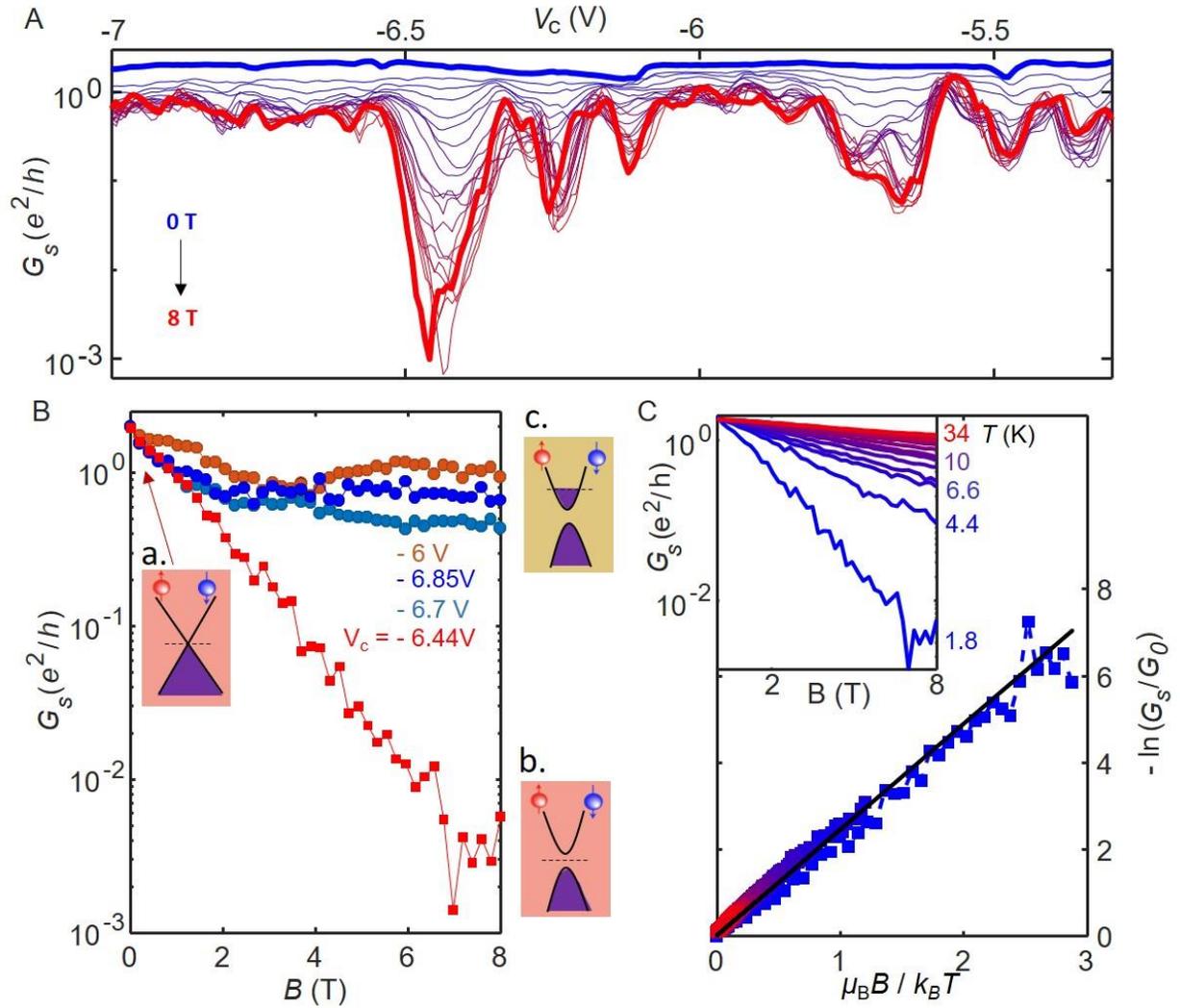

**Fig. 3. Time reversal symmetry breaking by a magnetic field and Zeeman-like gap at the Dirac point**. (A) The evolution of the edges conductance $G_s$ versus gate under the application of a perpendicular magnetic field, $B$ (from 0T, thick blue curve, to 8T, thick red curve) at 1.8K, for Device 1, 100 nm channel. (B) Traces of $G_s$ vs. $B$ for a few selected local gate voltages $V_c$ showing two types of behavior: saturation and non-saturation, associated with whether or not the Fermi energy is in the Zeeman gap, as depicted in the band schematics a. (linear bands at zero $B$, $E_F$ at Dirac point), b. (gapped bands at finite $B$, $E_F$ at Dirac point), and c. (gapped bands at finite $B$, $E_F$ away from Dirac point). (C) Inset: temperature dependence of $G_s$ vs. $B$ for the non-saturating curves ($V_c = -6.44V$). Main: All the curves in the inset collapse to a single trend in the normalized plot of $-\log(G_s/G_0)$ vs $\mu_B B/k_B T$. The black line is a linear fit. Additional inspection of the temperature- and magnetic-field dependence is shown in Fig. S9-11.

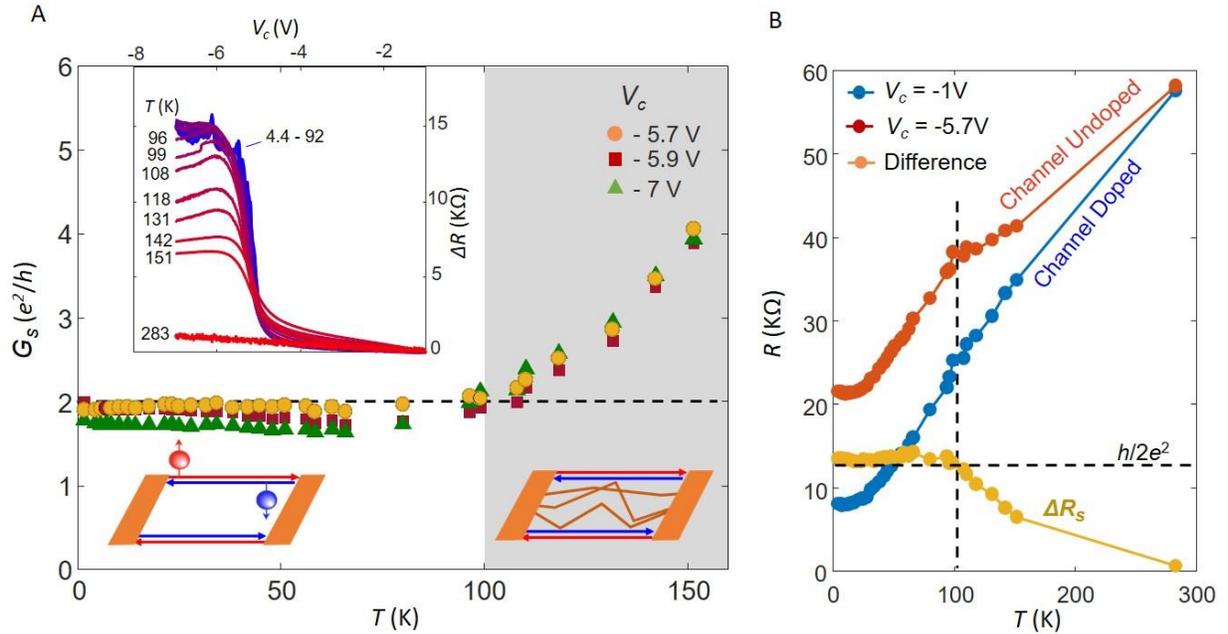

**Fig. 4. Quantum spin Hall effect up to 100 K**. (A) Temperature dependence of the edges conductance at a few representative gate voltages for 100 nm channel in Device 1. The conductance is dominated by the QSHE up to about 100K. The schematics depict the increase of the conductance due to onset of conduction from bulk states. Inset: gate dependence of $\Delta R$ at various temperatures. (B) Temperature dependence of the resistance of the whole flake (full length), when the Fermi energy in the local channel is in the doped ($V_c = -1$V, red) and undoped ($V_c = -5.7$V, blue) regimes, at $V_{tg} = 3.5$V. The difference between the curves yields the temperature dependent channel resistance $\Delta R_s$ (yellow). The vertical dashed line highlights the kink in the undoped regime at 100K, indicating the transition to the QSHE edge-dominated regime.

# Supplementary Materials for

# Observation of the Quantum Spin Hall Effect up to 100 Kelvin in a Monolayer Crystal

**Authors:** Sanfeng Wu[1,#,*], Valla Fatemi[1,#,*], Quinn D. Gibson[2], Kenji Watanabe[3], Takashi Taniguchi[3], Robert J. Cava[2], and Pablo Jarillo-Herrero[1,*]

[*]Correspondence to: swu02@mit.edu; vfatemi@mit.edu; pjarillo@mit.edu

**This PDF file includes:**

Materials and Methods
Supplementary Text
Figs. S1 to S11
Table S1

**Materials and Methods**

<u>General Fabrication Scheme</u>

The WTe$_2$ bulk crystals were grown as described in (*32*). The gates and electrodes of our devices are fabricated prior to the exfoliation of WTe$_2$. The detailed fabrication process is described below and illustrated in Fig. S1.

1. Creation of Gate Electrodes (Fig. S1(a-c))
    a. Dice Si$^{++}$/SiO$_2$ wafer into appropriately sized pieces, followed by spin-coating and baking a bilayer PMMA-based resist:
        i. 495A5, spun at 2000 rpm for 60 seconds, baked at 180C for 7 minutes
        ii. 950A5, spun at 3000 rpm for 60 seconds, baked at 180C for 3 minutes
    b. Electron beam lithography in an Elionix F125 system (acceleration voltage 125keV) to define gate electrodes with widths ranging from 50nm to 900nm.
    c. Develop resist in a cold water:IPA (1:3 by weight) mixture.
    d. Deposit Cr(3nm)/PdAu(30nm) in a thermal evaporator.
    e. Liftoff with successive baths of acetone and dichloromethane (DCM), followed by sonication in Remover PG, and final rinse in acetone and IPA.
    f. Heat clean at 300C for 3+ hours in forming gas (H$_2$ + Ar).
2. Transfer of bottom hBN (Fig. S1(d-f))
    a. Exfoliate hBN onto cleaned Si/SiO$_2$ wafer.
    b. Heat clean at 400C for 3+ hours in forming gas.
    c. AFM to ensure cleanliness of the flake.
    d. Pick up and transfer the flake onto the gates via standard dry transfer techniques using a polycarbonate/PDMS stamp.
    e. Remove the transfer polymer with chloroform.
    f. Heat clean at 300C for 3+ hours in forming gas.
3. Creation of Contact Electrodes (Fig. S1(g-i))
    a. Spin and bake of a PMMA-based bilayer resist recipe.
    b. Electron beam lithography to define contacts.
    c. Develop resist in a cold water:IPA (1:3 by weight) mixture.
    d. Deposit Ti(3nm)/PdAu(30nm) in a thermal evaporator. The height of the electrodes matches the local gates, to minimize the stressed (not fully encapsulated) region at the vicinity of the inner-most contacts.
    e. Liftoff with successive baths of acetone, DCM, and IPA.
    f. Tip clean surface of hBN with contact mode AFM.
    g. Mount chip into chip carrier and wire bond.
4. Transfer of WTe$_2$ and top gate electrode (Fig. S1(j,k))
    a. Prepare appropriate hBN and graphite pieces as per steps 2(a-c).
    b. WTe$_2$ flakes are exfoliated and identified in an argon glove box system with < 0.1 ppm of both O$_2$ and H$_2$O.
    c. Also in the glove box, pick up a global top hBN, then graphite top-gate electrode, then hBN to serve as top gate dielectric, and finally the target WTe$_2$ flake.
    d. Transfer entire stack onto pre-fabricated and pre-bonded gates and contacts.
    e. Remove transfer polymer with chloroform.
    f. Extract from glove box and immediately pump down in a cryostat.

Individual Device Details

Five devices are investigated in this study. Device 1 is discussed in depth in the main text, and images from its fabrication are shown in Fig. S1. Table S1 displays important parameters for each device.

**Table S1.**

Table of device parameters.

| Device # | Bottom hBN thickness | Top hBN thickness | Length between contacts | WTe$_2$ width |
|---|---|---|---|---|
| 1 | 13 nm | 9 nm | 7.2 μm | 1 μm |
| 2 | 16 nm | 9 nm | 4.0 μm | 3 μm |
| 3 | 10 nm | 11 nm | 4.0 μm | 4 μm |
| 4 | 19 nm | 10 nm | 7.2 μm | 7 μm |
| 5 | 15 nm | 9 nm | 2.7 μm | 4 μm |

Raman Analysis

We show a typical Raman spectrum taken from our exfoliated monolayer in Fig. S2. The spectrum is consistent with the literature(*12*, *16*, *33*), verifying the 1T' phase. The monolayer nature is also characterized by its transport behavior (*18*), as shown in Fig. 1B.

Measurement Details

Electronic transport measurements are conducted in a cryostat equipped with a superconducting magnet and a variable temperature $^3$He insert. The resistance is typically measured by applying a ~50μV low-frequency AC voltage source (~17Hz) using lock-in techniques. All the resistance data is measured under zero DC bias. DC bias almost has no effect on the resistance, as summarized in Fig. S3, indicating Ohmic contacts.

**Supplementary Text**

Extracting the Plateau Values

As mentioned in Fig. 1 of the main text, we performed a careful analysis to extract the value of the resistance plateaus. Generically, when we sweep the local gate voltage $V_c$ to negative values, we observe a resistance step in transition to the QSH plateau. This resistance step is measured as a function of the global top-gate voltage $V_{tg}$, as illustrated in Fig. S4. The onset of the step in $V_c$ changes with top gate voltage because the doping level local region is determined coordinately by both gates. In general, as the bulk becomes more highly doped by the top gate, the resistance step decreases and then converges, indicating the improvement of the contact between the doped bulk regions and the edge modes. We consider the converged value at the highest gate voltage as the extracted edge resistance $\varDelta R_s$, with an error given by the standard deviation of the step height at that top-gate voltage. We present figures displaying representative analysis of several cases (Fig. S4-7).

We also notice a particular case shown in Fig. S7 (the 500 nm-wide gate in Device 1), in which the resistance trace displays a second step appearing at even more negative $V_c$. This second step is a feature that appears for some of the longer channels (some hints of it are also visible in Fig. S4). A natural explanation is that the local gate dopes the channel into the valence band, transitioning from a n-edge-n device configuration of the first plateau to an n-edge-p-edge-n junction, as shown in the inset schematics. In this case, the second step would reflect the resistance change between the new n-edge-p-edge-n junction and n-edge-n junction. For example, the p-region may break the single edge mode into two edge modes in series, and may also scatter the carriers from one edge to the opposite. Therefore, a p region can reduce their transmission and result in an increase in resistance, as we observe. In our main analysis, we focus on the first step, which captures the resistance of just the edge mode with a length defined by the local gate.

## Bulk State Magnetic Field Dependence

The magnetoresistance in the highly doped bulk regime is small and approximately linear with field, as shown in Fig. S8. This magnetoresistance is largely temperature independent, and its weakness ($< 5\%$ change up to 8T) shows that the strong magnetic field dependence displayed in the edge regime does not originate from bulk states

## Saturation Conductance in High Magnetic Field

In the main text, it was noted that the edge conductance saturates to a finite value at high magnetic field for most gate voltages in the QSH regime. In Fig. S9(a), we show the gate dependence of the edge conductance for magnetic fields between 5 and 8 Tesla. The device has little change of behavior in this field range for most gate voltages away from Dirac point, again demonstrating the saturation. Fig. S9(b) shows characteristic magnetic-field dependences at the same gate voltages as in the main text (Fig. 4B), but on a linear scale. A histogram of all conductance extrema along gate voltage traces (determined algorithmically in MATLAB) at magnetic fields between 5 and 8 T produces the plot shown in Fig. S9(c). Theoretical predictions suggest that a single ballistic helical edge mode with a single charge puddle is expected to exhibit a 50% reduction of transmission at high magnetic fields(24), or similarly a dephasing charge puddle can effectively "break" a single QSH mode into two in series(30). Extending this to two edges and multiple puddles leads to a high-field conductance of $(1/(m+1) + 1/(n+1))\, e^2/h$ where $m$ and $n$ are integers that count the number of charge puddles coupled to each edge. The expected saturated values (0.5, 0.66, 0.83 and 1) $e^2/h$ for several cases of $m$ and $n$ are indicated in the plots. Our data is suggestive of the prediction, with the number of charge puddles varying with the local gate voltage but usually very few ($< 3$ per edge). Further investigation in both experiment and theory are necessary to fully understand the mechanism.

## In-Plane Magnetic Field

We also find that in-plane magnetic fields degrade the edge conductance. In Fig. S10, we show the edge conductance as a function of local gate voltage for a 100nm-wide gate in Device 2. In this case, we don't see a clear Dirac point but rather a more uniform degradation of the edge conductance for all gate voltages. The absence of the Dirac point in this case requires future work to understand. Generally, the observation of a gap opening at the Dirac point requires a magnetic field non-parallel to the spin polarization axis of the edge modes as well as nearly identical conditions for both edges, which may include offset density, disorder strength, crystallographic orientation, and edge termination. Moreover, crystallographic orientation and edge termination may influence whether a Dirac point in the edge state exists inside the bulk gap. We currently do not have the fine control over these parameters to engage in a targeted study of this physics. Nonetheless, our observations encourage further studies of the Dirac point in monolayer TIs, both experimentally and theoretically.

Temperature Dependence of the Edge Resistance in High Magnetic Field

As shown in Fig. 3(c) of the main text, the edge resistance has temperature and magnetic field dependences that suggest a Zeeman-like gap in a narrow gate voltage range. All of the high-resistance features that exist at low temperature are strongly suppressed with increasing temperature. In Fig. S11 we show the gate voltage dependence as a function of magnetic field for four representative temperatures, observing that sharp and tall features become broad and shallow as temperature is raised.

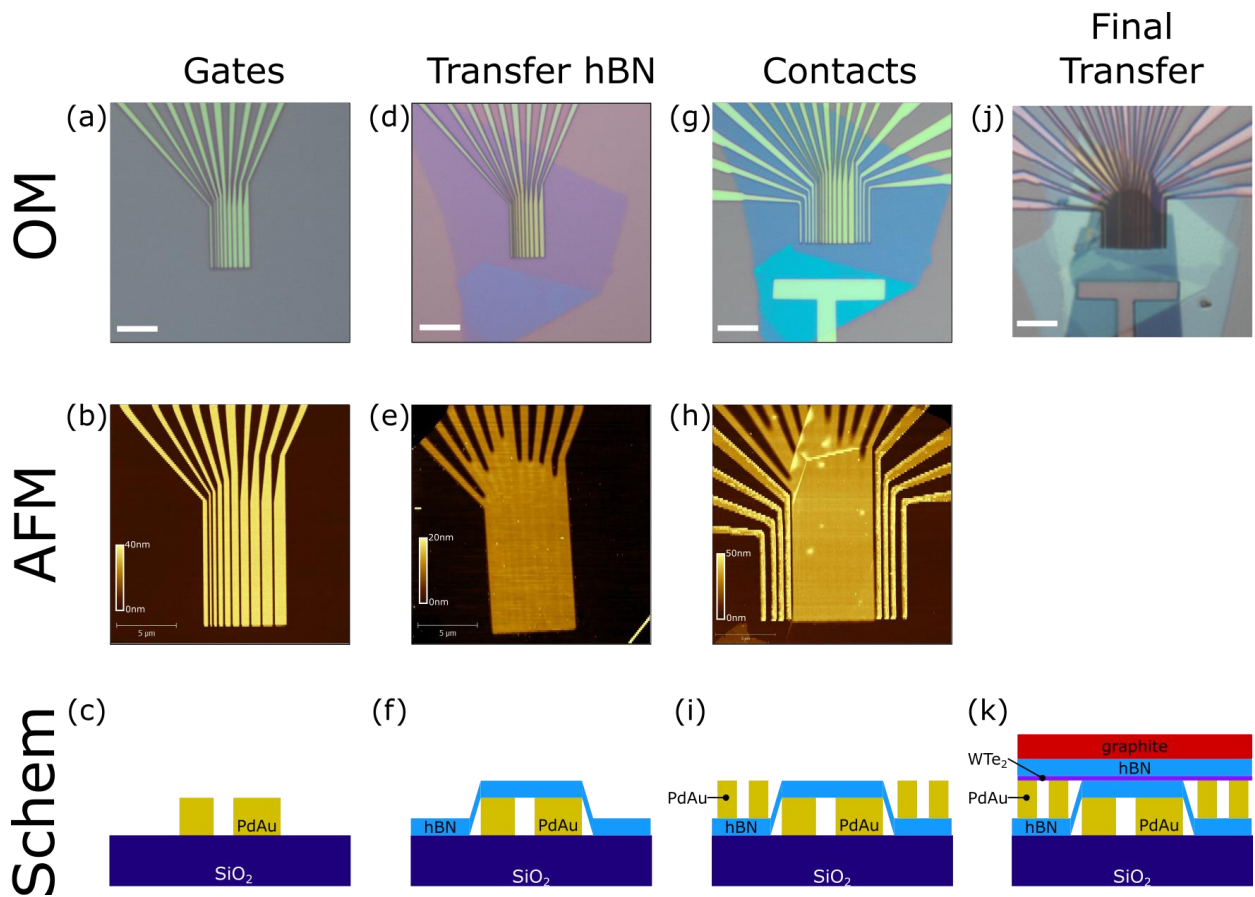

**Fig. S1 Images of fabrication steps.** Optical microscopy (OM) images (upper row) and atomic force microscopy (AFM) images (middle row) taken at each key stage of the fabrication process. The images in the lower row are cross-sectional schematics to illustrate the structure.

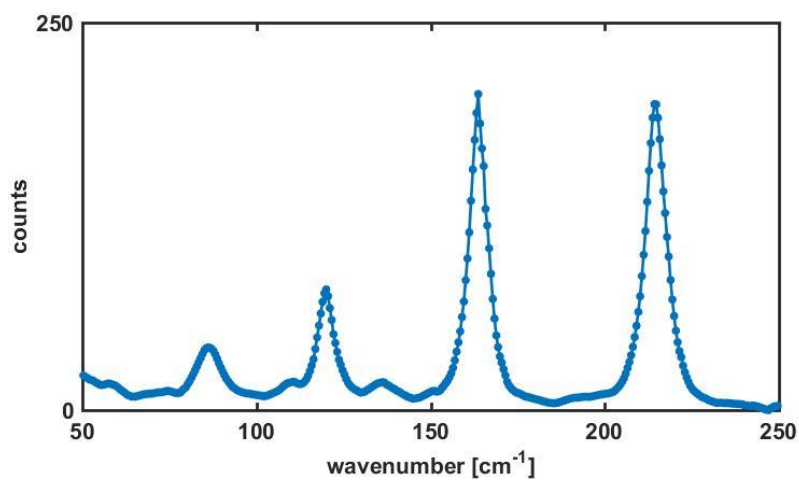

**Fig. S2 Raman spectroscopy of monolayer WTe$_2$.** Raman spectrum of an exfoliated monolayer of WTe$_2$ taken at room temperature with a 532nm excitation laser, consistent with the 1T' phase of monolayer WTe$_2$.

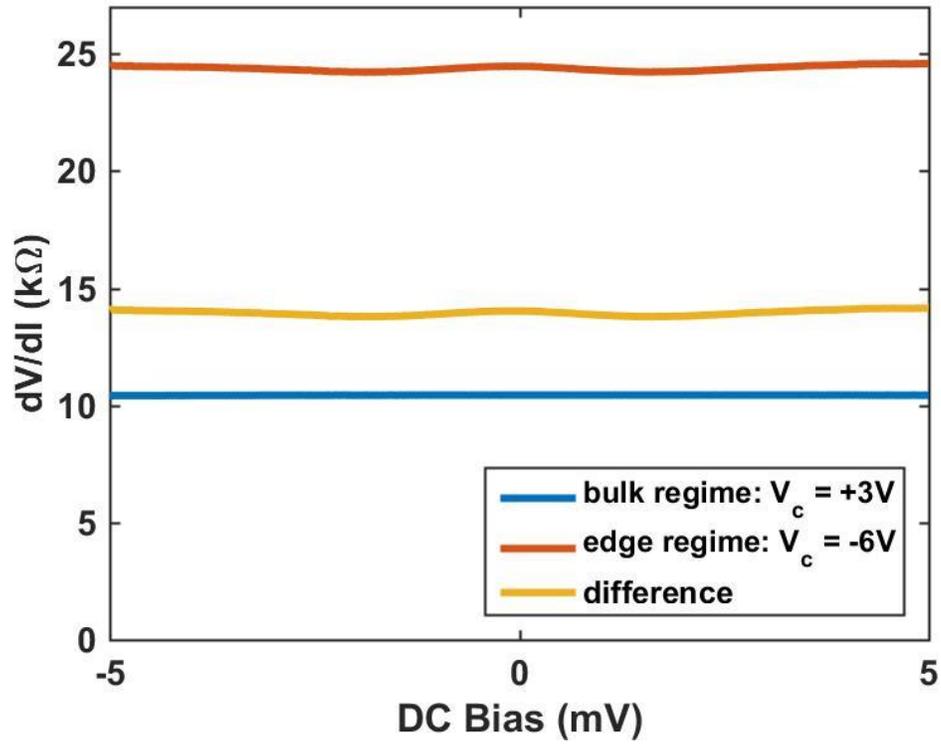

**Fig. S3 Ohmic contacts.** dV/dI as a function of DC bias for $V_{tg}$ = 3.5V on device 1 measured in four-terminal configuration at 4K. Blue and orange correspond to the bulk regime ($V_c$ = 3V) and edge regime ($V_c$ = -6V), respectively. The difference is in yellow. The flat dV/dI indicates linear ohmic contacts.

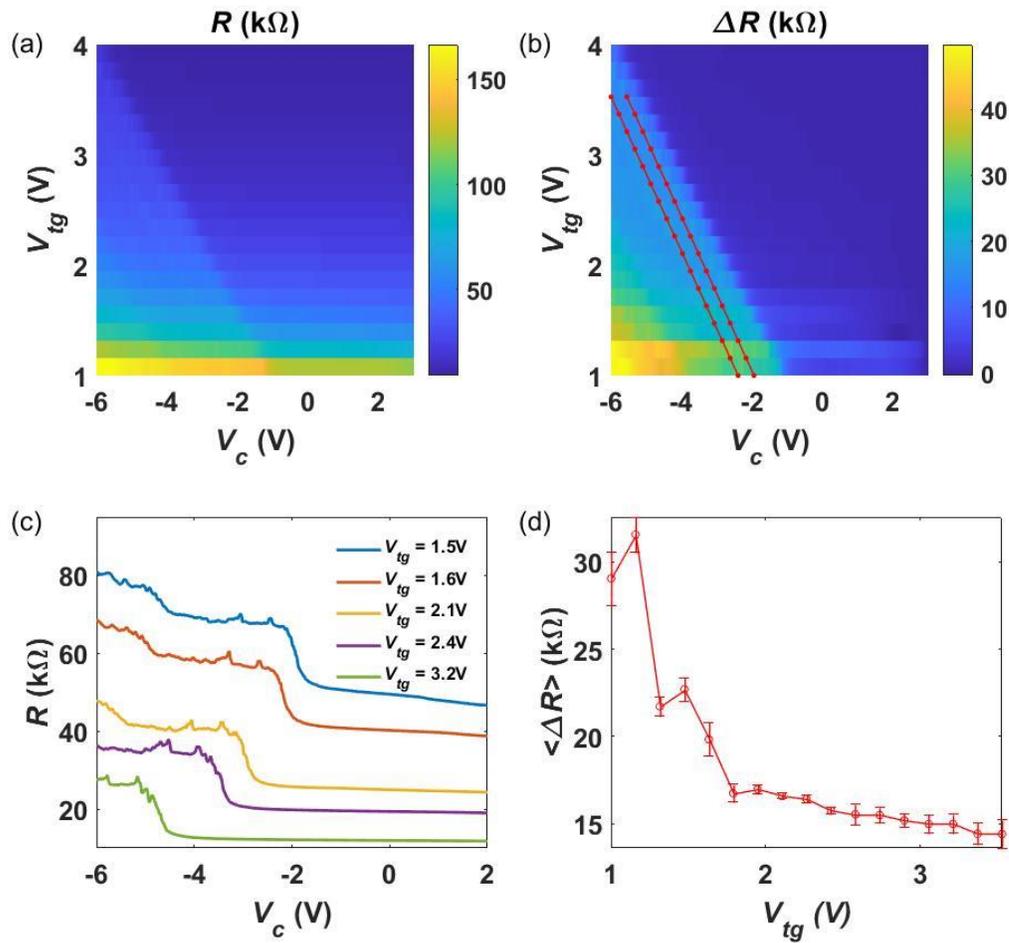

**Fig. S4 Analysis to extract the edge resistance – Device 1, 100nm gate, 4 K.** (a) Total resistance as a function of top gate and local gate voltage ($V_{tg}$ and $V_c$, respectively), for device 1 with the 100nm-wide local gate. (b) The same data as (a) with the resistance at $V_c = 3V$ subtracted away, which we designate $\Delta R$. The space between the red dotted lines indicates the range of $\Delta R$ on the step that is averaged to extract the step height. (c) Specific traces of total resistance vs $V_c$ showing how the traces change progressively with increasing top-gate voltage. (see SI text and Fig. S7 for explanation on signatures of second step). (d) A plot of the step resistance as a function of $V_{tg}$, showing the saturation at large $V_{tg}$. The error bars indicate a full standard deviation from the mean $\Delta R$ in the range given by the red dots in (b).

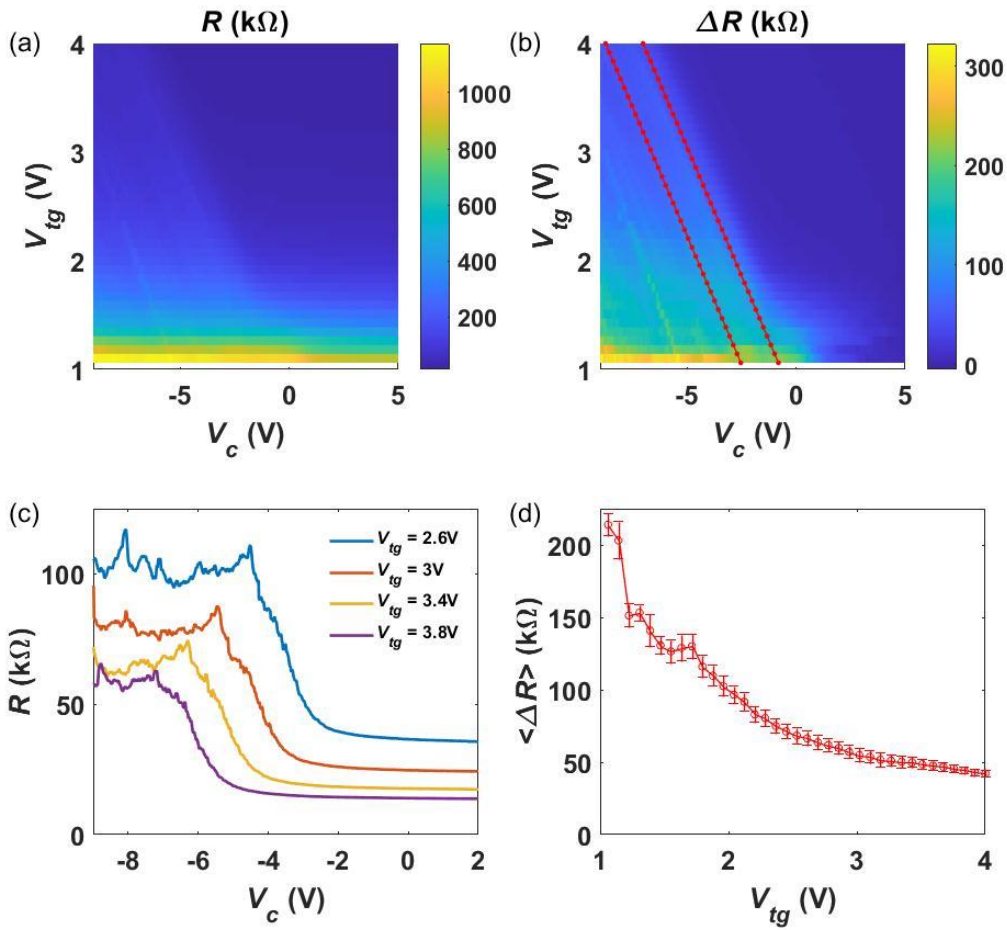

**Fig. S5 Analysis to extract the edge resistance – Device 4, 400nm gate, 4 K.** Same as Fig.S4, but for the 400nm-wide gate in Device 4. The resistance of this long channel saturates to a resistance well above what is measured in the short-channel limit. In general, we have observed increased resistance with increasing length in long channel devices (Fig. 2). This increase of resistance can be understood by considering dephasing scatterers at the edges. The helical edge mode is equilibrated at the dephasing sites, which effectively break the long edge mode into multiple in series, resulting in a trend of increasing total resistance. Device-to-device variation for the same channel length can thus be attributed to differing disorder realizations for the different devices.

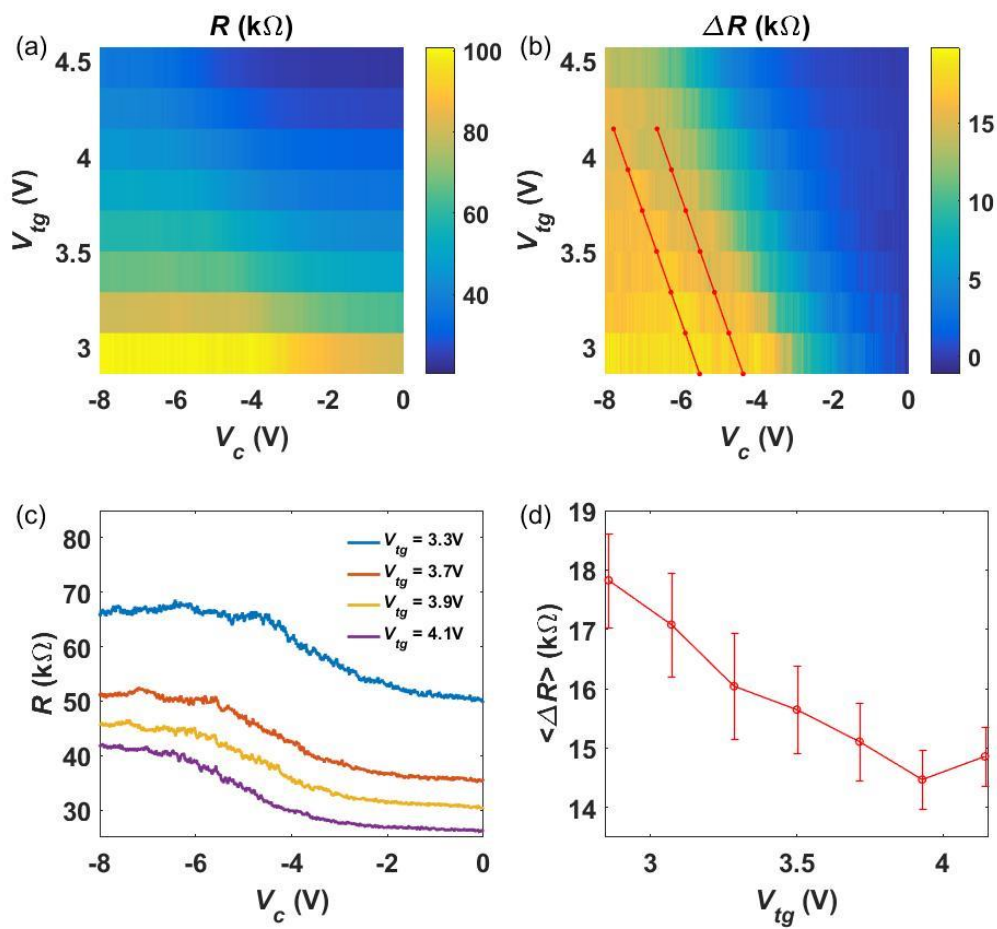

**Fig. S6 Analysis to extract the edge resistance – Device 2, 100 nm gate, 4 K.** Same as Fig. S4, but for the 100nm-wide gate in Device 2, which shows a similar converged resistance step value as in Device 1.

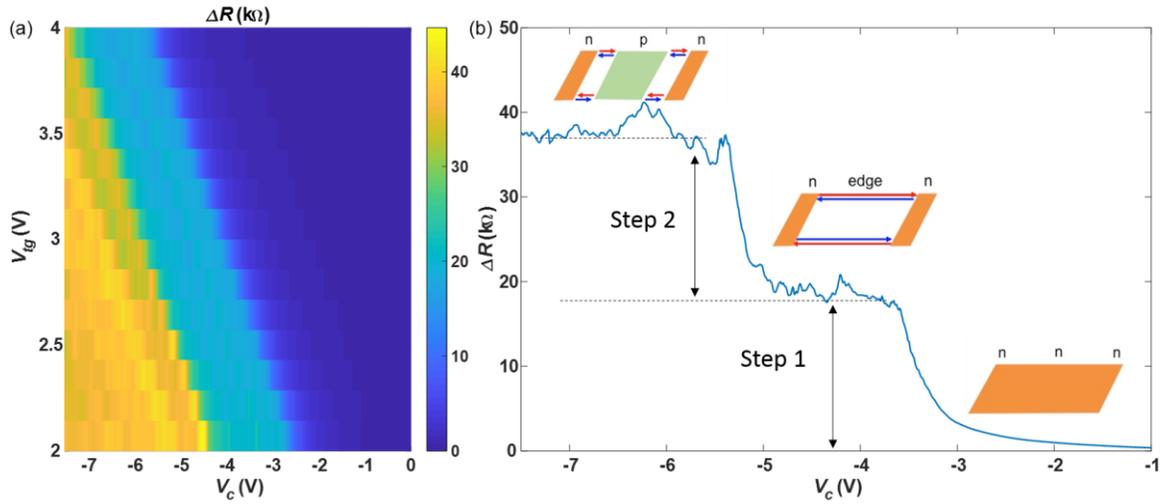

**Fig. S7 Channel resistance with a second step.** (a) $\Delta R$ as a function of both gate voltages and (b) a representative line-cut of $\Delta R$ vs. $V_c$ at $V_{tg} = 2.6$ V for the 500nm-wide local gate in Device 1 at 4 K. The second step exhibited here appears especially for a few of our longer channels. It could be related to the creation of an n-edge-p-edge-n junction as indicated by the schematics in panel (b). See supplementary text for details. The height of the first step is extracted as the channel resistance of the edge mode with length defined by the local gate.

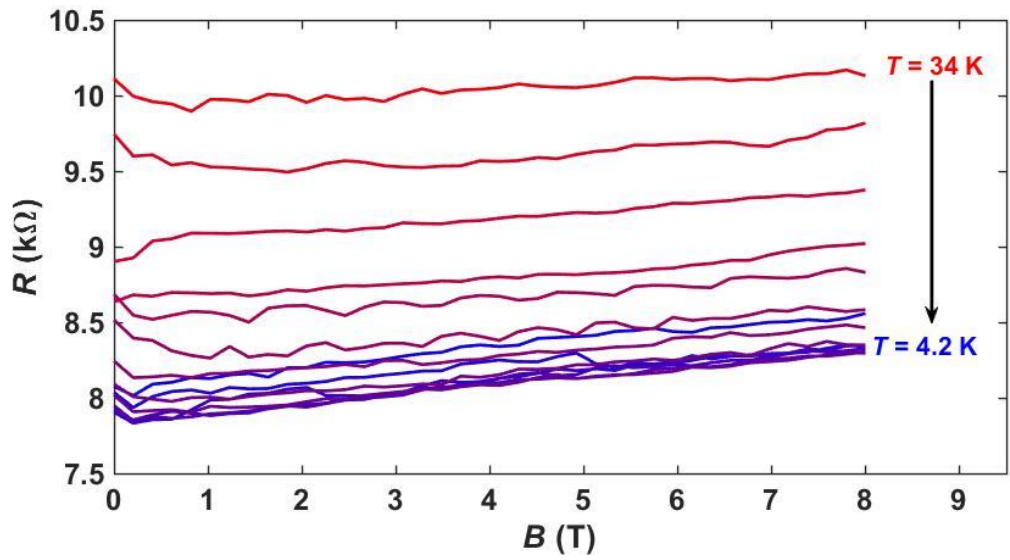

**Fig. S8 Weak magnetoresistance in the bulk-doped regime.** Magnetic field dependence of the resistance in the bulk-doped regime ($V_{tg}$ = +3.5 V) at different temperatures. The weak magnetoresistance here shows that the strong magnetoresistance observed in the edge regime is unrelated to bulk states.

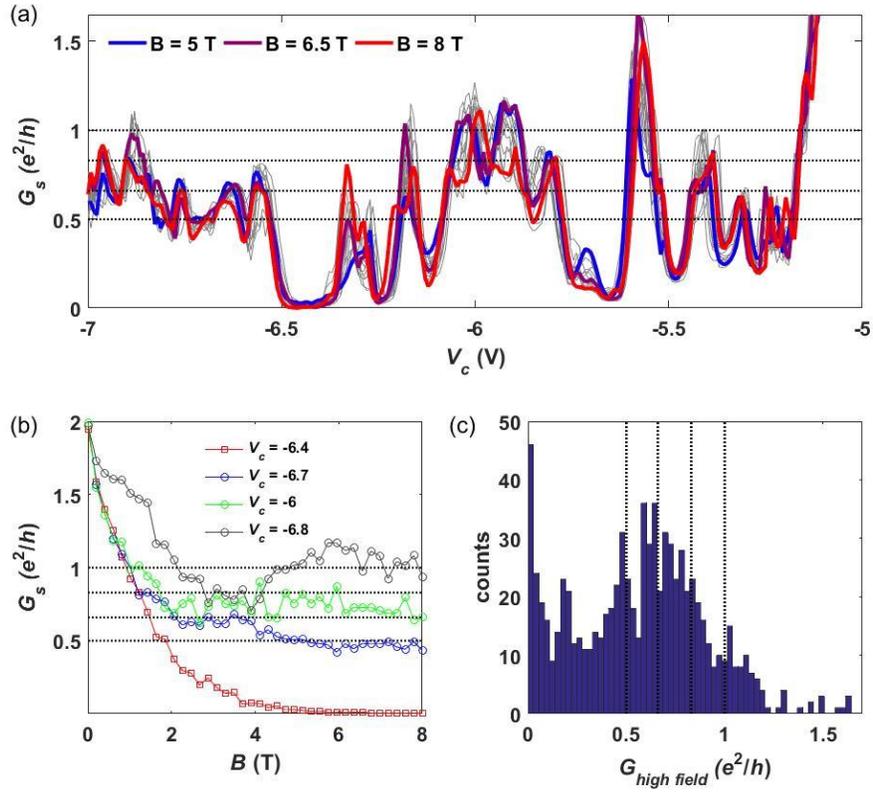

**Fig. S9 Saturation conductance of the edge modes at high magnetic fields.** (a) Magnetic field dependence of the edge conductance from $B = 5T$ to 8T, with three specific traces highlighted and all others in grey. The dotted lines indicate values (0.5, 0.66, 0.83, 1) expected from scenarios based on charge puddles (see supplemental text for details). (b) A few specific traces of the edge conductance as a function of magnetic field. In red is the exponential behavior near the Dirac point, and the other three traces show typical saturation behavior. Dotted lines are the same as in (a). (c) A histogram of all local conductance extrema along gate voltage traces ($G_{high\,field}$) from 5T to 8T, with dotted lines again highlighting the same values (see supplementary text for details).

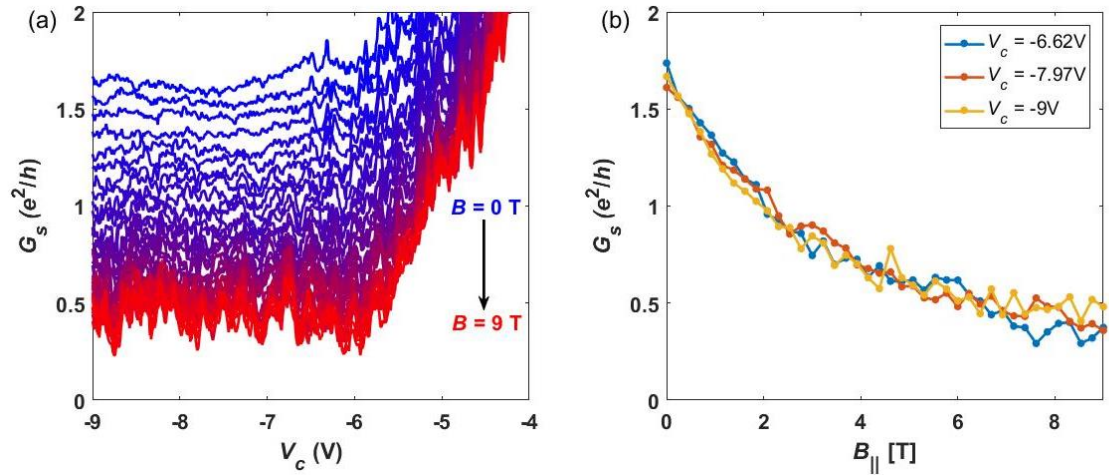

**Fig. S10 In-plane magnetoresistance.** (a) Edge conductance as a function of local gate voltage $V_c$ for different in-plane, roughly perpendicular-to-edge, magnetic field strengths at 4 K for the 100nm-wide gate of Device 2. (b) Line traces of resistance as a function of magnetic field for selected $V_c$. The conductance is clearly suppressed under in-plane magnetic fields. This device does not have a distinct Dirac point, but instead sees a uniform decrease in conductance, possibly due to a different edge configuration or non-uniform disorder strength at the edges compared to Device 1.

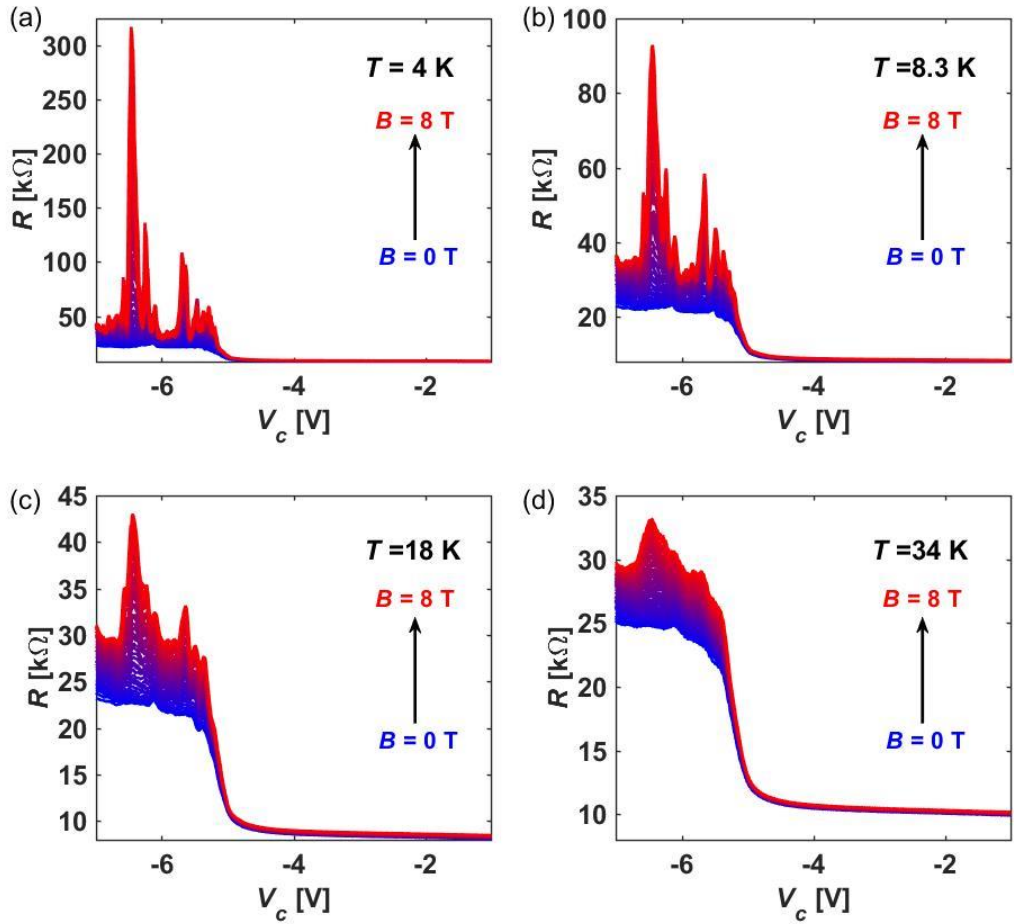

**Fig. S11 Magnetic field effect at different temperatures.** Raw data for the temperature dependent magnetoresistance measurements for Device 1, 100 nm local channel. Resistance vs. local gate voltage $V_c$ subjected to different magnetic fields for a representative set of temperatures: (a) 4 K, (b) 8.3 K, (c) 18K, and (d) 34 K.